 \documentclass[3p,times]{elsarticle}

\usepackage{amssymb}
\usepackage{amsmath, bbm}
\usepackage[figuresright]{rotating}
\usepackage{capt-of}
\usepackage{multirow}

\usepackage{epsfig}
\def\be{\begin{equation}}
\def\ee{\end{equation}}
\def\ba{\begin{eqnarray}}
\def\ea{\end{eqnarray}}

\newcommand{\ppbar}{{\bar p p}}
\newcommand{\eebar}{{e^+e^-}}
\newcommand{\nnbar}{{\bar N  N}}
\newcommand{\ccc}{\cdot\cdot\cdot}

\begin{document}

\begin{frontmatter}

\title{The electromagnetic form factors of the proton in the
timelike region}

 \author[J]{J. Haidenbauer}
 \author[J]{X.-W. Kang}
 \author[B,J]{U.-G. Mei{\ss}ner}
 \address[J]{Institute for Advanced Simulation, Institut f{\"u}r Kernphysik and
 J\"ulich Center for Hadron Physics, Forschungszentrum J{\"u}lich, D-52425 J{\"u}lich, Germany}
 \address[B]{Helmholtz Institut f\"ur Strahlen- und Kernphysik and Bethe Center
  for Theoretical Physics, Universit\"at Bonn, D-53115 Bonn, Germany}

\begin{abstract}
The reactions $\ppbar \to \eebar$ and $\eebar \to \ppbar$
are analyzed in the near-threshold region. Specific emphasis 
is put on the role played by the interaction in the initial- or
final antinucleon-nucleon ($\nnbar$) state which is taken into account rigorously.
For that purpose a recently published $\nnbar$ 
potential derived within chiral effective field theory and fitted to results 
of a new partial-wave analysis of $\ppbar$ scattering data is employed.
Our results provide strong support for the conjecture 
that the pronounced energy dependence of the
$\eebar \leftrightarrow \ppbar$ cross section, seen in pertinent 
experiments, is primarily due to the $\ppbar$ interaction. 
Predictions for the proton electromagnetic form factors $G_E$ and $G_M$ 
in the timelike region, close to the $\nnbar$ threshold, and for 
spin-dependent observables are presented. 
The steep rise of the effective form factor for energies close to
the $\ppbar$ threshold is explained solely in terms of the 
$\ppbar$ interaction. The corresponding experimental information 
is quantitatively described by our calculation. 
\end{abstract}
\end{frontmatter}

\section{Introduction}

The electromagnetic form factors (EMFFs) of the proton and the neutron play an 
important role in our understanding of the nucleon structure. Experimental and 
theoretical studies of these quantities in the spacelike region, i.e. 
in electron-proton scattering, started already more than half a 
century ago. Over the last decades there is also an increased interest
in their properties in the timelike region, accessible in
the reactions $\ppbar \to \eebar$ and $\eebar \to \ppbar$, 
as witnessed by various publications 
\cite{Bijker,Belushkin,Dubnickova,Melo,Yan,Kuraev,Simonov,Lomon}
and a recent extensive review article \cite{Denig}. 
In particular, the observation of a strong energy dependence of the proton 
EMFFs close to the $\ppbar$ threshold, 
i.e. at momentum transfers $q^2\simeq (2 M_p)^2$, has attracted quite some attention. 
This behavior was first reported by the PS170 collaboration~\cite{Bardin}, 
and detected in a measurement of the $\ppbar \to \eebar$ reaction cross 
section at LEAR. 
In recent years the BaBar collaboration has measured the cross section for 
the time-reversed process $\eebar \to \ppbar$ \cite{Aubert0,Lees}.  
Their data are of similar precision as those from the PS170
collaboration and cover also energies very close to the $\ppbar$ threshold.
The form factor deduced from those data substantiates the 
finding of the PS170 collaboration. 

A strong dependence of the proton EMFFs on the momentum transfer simply
reflects the fact that the underlying (measured) $\eebar \to \ppbar$ cross 
section shows a significant enhancement near the $\ppbar$ threshold. 
Such near-threshold enhancements were also reported in entirely different 
reactions involving the $\ppbar$ system, 
for example, in the $\psi(3686) {\to}\gamma \ppbar$ \cite{BES12}
and the $B^+ \to \ppbar K^+$ \cite{Aubert:2005gw}
decays, and in particular in 
the radiative decay $J/\psi \to \gamma \ppbar$ \cite{BES12,Bai}. 
For the latter case several explanations have been put forth, including 
scenarios that invoke $N{\bar N}$ bound states or so far unobserved
meson resonances. However, it was also shown that a conventional but
plausible interpretation of the data can be given simply in terms of the final-state 
interaction (FSI) between the produced proton and antiproton 
\cite{Sibirtsev1,Loiseau,Kerbikov,Bugg1,Zou}. 
Specifically, calculations of our group,
utilizing the J\"ulich $N{\bar N}$ model \cite{Hippchen,Hippchen2,Mull} 
and performed within the Watson-Migdal approach \cite{WM,WM1}, 
could reproduce the mass dependence of the $\ppbar$ spectrum close to 
the threshold by the $S$-wave $\ppbar$ FSI for various decays
\cite{Sibirtsev1,Haidenbauer2006,Haidenbauer2012}. 
 
The success of those investigations suggests that the same effects, namely
the FSI between proton and antiproton, could be also responsible for the
near-threshold enhancement in the $\eebar \to \ppbar$ cross
section and, accordingly, for the strong $q^2$ dependence of the 
proton EMFF in the timelike region near $q^2\approx (2 M_p)^2$. 
Indeed, a few years ago we have studied the energy dependence of the 
$\eebar \leftrightarrow \ppbar$ cross section close to threshold,
within the Watson-Migdal approach \cite{Sibirtsev}.
We could show that the near-threshold enhancement in the
$\eebar \to \ppbar$ cross section can be explained qualitatively 
by $\ppbar$ FSI effects in the $^3S_1$ partial wave as generated by 
the J\"ulich nucleon-antinucleon model \cite{Hippchen}.
Similar results were also reported by other authors based on
somewhat different approaches and employing other $\nnbar$ interactions
\cite{Dmitriev07,Chen08,Chen10,Dalkarov,Dmitriev11,Dmitriev13,Dmitriev13a}.
 
The present study of the proton EMFF in the timelike region aims at 
an improvement of our earlier work \cite{Sibirtsev} in various aspects:  
First and foremost the new calculation of the $\eebar \leftrightarrow \ppbar$ 
transition is based on a refined and formally exact treatment of the 
effects from the $\nnbar$ interaction in the initial or final state. 
Second, we take into account the coupling between the $^3S_1$ and $^3D_1$ 
partial waves. In the commonly adopted one-photon approximation 
these are the only two partial waves that can contribute. The inclusion
of the $^3D_1$ state allows us to extend the energy range of our
study. Furthermore, it enables us to obtain non-trivial results for angular distributions 
and compare those to available data, and we can make concrete predictions for 
(not yet measured) spin-dependent observables. 
Finally, in the meantime results of a new partial-wave analysis (PWA) of $\ppbar$ 
scattering data have been published \cite{Zhou2012}. Based on that work an 
$\nnbar$ potential has been constructed by us \cite{Kang}, in the framework of 
chiral effective field theory (EFT), that reproduces the amplitudes determined 
in the PWA very well up to laboratory energies of $T_{lab} \approx 200-250$ MeV. 
This potential will be now employed for the final-state interaction, besides the 
phenomenological $\nnbar$ model of the J\"ulich group \cite{Hippchen} used 
in our earlier work \cite{Sibirtsev}. 

The paper is structured in the following way: In the subsequent section
we summarize the formalism. Specifically, we provide details about how the
$\ppbar$ FSI is included in our calculation. 
In Sect. 3 we compare our results with measured integrated and differential 
cross sections for the reactions $\eebar\to \ppbar$ and $\ppbar \to \eebar$ 
in the region near the $\ppbar$ threshold. Furthermore, we provide 
predictions for spin-dependent observables for which so far there is no
experimental information. Finally, we present results for the EMFFs
$G_E$ and $G_M$, for their ratio as well as for the relative
phase. 
The paper closes with a summary. 

%%%%%%%%%%%%%%%%%%%%%%%%%%%%%%%%%%%%%%%%%%%%%%%%%%%%%%%%%%%%%%%%%%%%%%%%%%%%%%%%%
\section{Formalism}
\label{sec:form}

We adopt the standard conventions so that the differential cross section
for the reaction $\eebar \to \ppbar$ is given by \cite{Denig} 
\be
\frac{d\sigma}{d\Omega} = \frac{\alpha^2\beta}{4 s}
~C_p(s)~
\left[\left| G_M(s) \right|^2 (1+{\rm cos}^2\theta) +
\frac{4M_p^2}{s} \left| G_E(s) \right|^2 {\rm sin}^2\theta \right]~{\rm .}
\label{eq:diff}
\ee
Here, $\alpha = 1/137.036$ is the fine-structure constant and 
$\beta=k_p/k_e$ a phase-space factor, where $k_p$ and $k_e$ are the
center-of-mass three-momenta in the $\ppbar$ and $\eebar$ systems, respectively,
related to the total energy via $\sqrt{s} = 2\sqrt{M_p^2+k_p^2} = 2\sqrt{m_e^2+k_e^2}$. 
Further, $m_e \, (M_p)$ is the electron (proton) mass.
The $S$-wave Sommerfeld-Gamow factor $C_p(s)$ is given by $C_p = y/(1-e^{-y})$ 
with $y = \pi \alpha M_p /k_p$. 
$G_E$ and $G_M$ are the electric and magnetic form factors, respectively. 
The cross section as written in Eq.~({\ref{eq:diff}) results from the one-photon 
exchange approximation and by setting the electron mass $m_e$ to zero
(in that case $\beta = 2k_p/\sqrt{s}$). We will restrict
ourselves throughout this work to the one-photon exchange so that the total 
angular momentum is fixed to $J=1$ and the $\eebar$ and $\nnbar$ system can be only in 
the partial waves $^3S_1$ and $^3D_1$. We use the standard spectral notation
$^{(2S+1)}L_J$, where $S$ is the total spin and $L$ the orbital angular momentum.  
Let us mention that there are indications that two-photon exchange contributions are 
important in the spacelike region and can account for the discrepancy between the 
form factor values extracted from polarization data and from Rosenbluth separation
of cross section data \cite{GV03,BMT03,CABCV04,RT04,Belushkin:2007zv,Arrington:2011dn}. 
Their importance in the timelike region is less clear, see for example 
Refs.~\cite{GaT05,GaT06}.

The integrated reaction cross section is readily found to be 
\be
\sigma_{\eebar \to \ppbar} = \frac{4 \pi \alpha^2 \beta}{3s} ~C_p(s)~
\left [ \left| G_M(s) \right|^2 + \frac{2M_p^2}{s} \left| G_E(s) \right|^2 \right ]~{\rm .}
\label{eq:tot}
\ee

Another quantity used in various analyses is the effective proton form factor $G_{\rm eff}$
which is defined by 
\be
|G_{\rm eff} (s)|=\sqrt{\sigma_{\eebar\rightarrow \ppbar} (s)\over {4\pi\alpha^2
\beta \over 3s} ~C_p(s)   \left [1
+\frac{2M^2_p}{s}\right ]} \ . 
\label{eq:Geff}
\ee

In the helicity basis, the amplitudes for the reaction $\eebar \to \ppbar$ for one-photon exchange
are given by \cite{Buttimore31,Buttimore33}  
\begin{eqnarray}
  \label{hel1}
\phi_1 = \langle ++|F|++\rangle &=&
-\frac{2\,m_e\,M_p\alpha}{s}\,\cos\theta\,G_E = \langle ++|F|--\rangle = \phi_2 ~, \nonumber
\\\nonumber
 \\\nonumber
\phi_3 = \langle +-|F|+-\rangle &=&
-\frac{\alpha}{2}\,\left( 1 + \cos\theta \right)\,
G_M ~,\nonumber
 \\\nonumber
\\
\phi_4 = \langle +-|F|-+\rangle &=&
-\frac{\alpha}{2}\,\left( 1 - \cos\theta \right)\,
G_M ~,
\\\nonumber
 \\\nonumber
\phi_5 = \langle ++|F|+-\rangle &=&
\frac{M_p\alpha}{\sqrt{s}}\,\sin\theta\, G_E = -\langle ++|F|-+\rangle = -\phi_7 ~,
\nonumber
 \\\nonumber
 \\\nonumber
\phi_6 = \langle +-|F|++\rangle &=&
-\frac{m_e\alpha}{\sqrt{s}}\,\sin\theta\,G_M = -\langle -+|F|++\rangle = -\phi_8 \ . 
\label{eq:Hel}
 \end{eqnarray}
For convenience we include the electron mass explicitly here and in the 
formulae below and also in our numerical calculation. 
In terms of those amplitudes the differential cross section is given by 
\be
\frac{d\sigma}{d\Omega} = \frac{1}{2s}\,\beta \ C_p \sum_{i=1}^8 |\phi_i|^2~,
\label{eq:Sum}
\ee
which reduces to the result in Eq.~(\ref{eq:diff}) for $m_e \to 0$. 
Note that the amplitudes for the inverse reaction $\ppbar \to \eebar$ are
given by the same expressions but with the obvious replacements 
$\phi_5 \to -\phi_6$ and $\phi_6 \to -\phi_5$.   

In order to implement the FSI we perform a partial wave projection of the 
$\eebar \to \ppbar$ amplitudes and switch from the helicity basis to the 
more convenient $JLS$ representation. The corresponding formalism is documented 
in various publications in the literature. We follow here the procedure described
in detail in the Appendices B and C of Ref.~\cite{Holzenkamp}. 
Then we end up with four amplitudes, corresponding to the 
coupling between the $\eebar$ and the $\ppbar$ systems and 
the coupled $^3S_1$-$^3D_1$ partial waves. 
We can write these in the form $F_{L\,L'}$, where $L' (L)=0,2$
characterizes the orbital angular momentum in the initial (final)
state. The explicit expressions for the reaction $\eebar \to \nnbar$
are
\ba
\nonumber
F^{\mu,\nu}_{2\,2} &=& -\frac{2\alpha}{9} \left[ G_M - \frac{2 M_p}{\sqrt{s}} G_E\right]
\left[ 1 - \frac{2 m_e}{\sqrt{s}} \right]~, \\
\nonumber
F^{\mu,\nu}_{0\,0} &=& -\frac{4\alpha}{9} \left[ G_M + \frac{M_p}{\sqrt{s}} G_E\right]
\left[ 1 + \frac{m_e}{\sqrt{s}} \right]~, \\
\nonumber
F^{\mu,\nu}_{0\,2} &=& -\frac{2\sqrt{2}\alpha}{9} \left[ G_M + \frac{M_p}{\sqrt{s}} G_E\right]
\left[ 1 - \frac{2 m_e}{\sqrt{s}} \right]~, \\
F^{\mu,\nu}_{2\,0} &=& -\frac{2\sqrt{2}\alpha}{9} \left[ G_M - \frac{2 M_p}{\sqrt{s}} G_E\right]
\left[ 1 + \frac{m_e}{\sqrt{s}} \right] \ . 
\label{eq:pwa} 
\ea
For reasons of clarity we include in Eq.~(\ref{eq:pwa}) and in the next few lines 
superscripts for the channels ($\nu = \eebar$ and $\mu = \ppbar$), but we will omit 
them again later in order to simplify the notation. Time reversal invariance requires 
that $F^{\mu,\nu}_{L\,L'}(p,p') = F^{\nu,\mu}_{L'\,L}(p',p)$
so that for the reaction $\ppbar \to \eebar$ the amplitudes 
$F_{0\,2}$ and $F_{2\,0}$ are interchanged. 
 
\begin{figure}[t]
\begin{center}
\includegraphics[height=180mm]{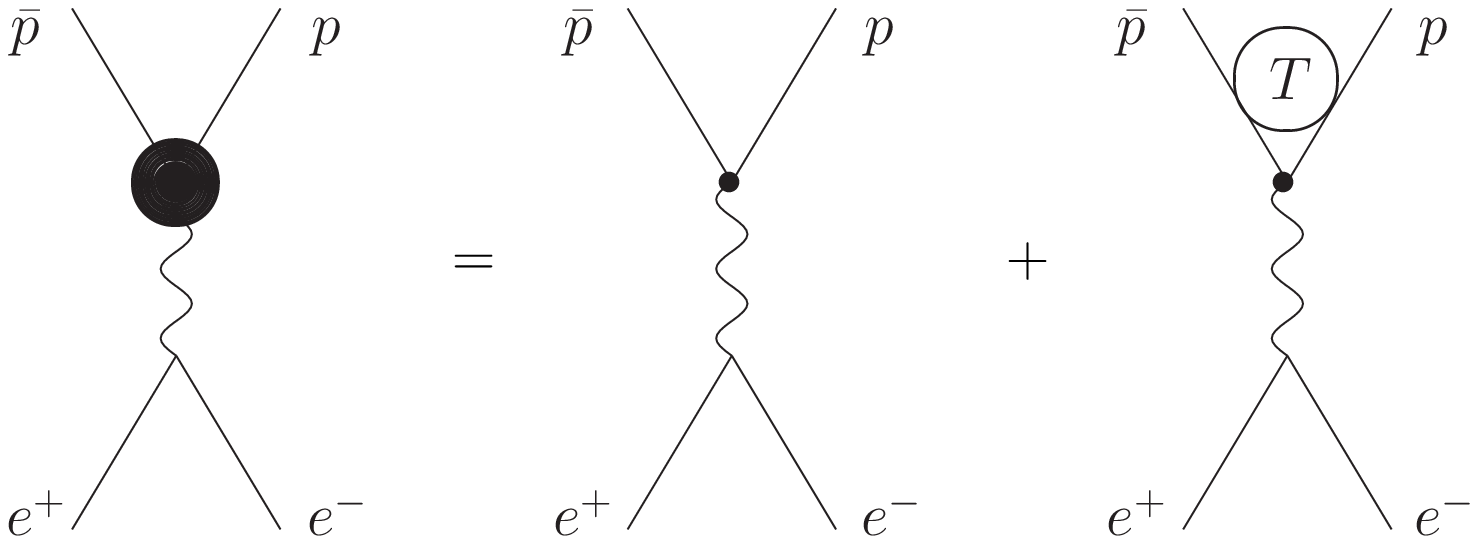}
\end{center}
\vskip -13.0cm
\caption{Graphic representation of our treatment of the
reaction $\eebar \to \ppbar$. The small (large) filled circle symbolizes the
bare (dressed) $\nnbar\gamma$ vertex while $T$ stands for the 
$\nnbar$ scattering amplitude. 
}
\label{Diag}
\end{figure}

It is obvious from Eq.~(\ref{eq:pwa}) that the amplitude $F^{\mu,\nu}_{L\,L'}$ can
be written as a product of factors, which is simply a consequence of
the one-photon exchange which amounts to an $s$-channel pole diagram in
the reactions $\eebar \leftrightarrow \ppbar$. The factors correspond to
the $\eebar\gamma$ and $\ppbar\gamma$ vertices, respectively, and 
reflect whether the coupling occurs in an $S$ or $D$ wave.
Thus, we can write the amplitude in the form ($L,\,L'=0,2$)
\ba
\nonumber
F^{\mu,\nu}_{L\,L'} &=& -\frac{4\alpha}{9} \ f^{\mu}_L \ f^{\nu}_{L'}, \ \ \  {\rm with}  \\
f^{\mu}_0 &=& \left(G_M + \frac{M_p}{\sqrt{s}} G_E\right), \
f^{\mu}_2= \frac{1}{\sqrt{2}}  \left(G_M - \frac{2 M_p}{\sqrt{s}} G_E\right)~, 
\label{FOFA} 
\ea
and similar expressions for $f^{\nu}_L$, the vertex functions of the 
$\eebar$ pair. The FSI effects due to the $\ppbar$ interaction influence only the 
$\ppbar$ vertex and that means only $f^{\mu}_L$ (simply denoted by $f_L$ in the following), 
see Fig.~\ref{Diag}.
These effects can be calculated rigorously and within our formalism they amount 
to evaluating the equation 
\begin{eqnarray}
&&f_{L'}(k;E_k)=f^0_{L'}(k)+
%\nonumber\\&&
\sum_{L}\int_0^\infty \frac{dpp^2}{(2\pi)^3} \, f^0_{L}(p)
\frac{1}{2E_k-2E_p+i0^+}T_{LL'}(p,k;E_k) \ , 
\label{FSI}
\end{eqnarray}
where the first term on the right-hand side, the so-called
Born term, represents the bare $\nnbar$ production vertex
$f^0_L$ and the integral provides the dressing of this vertex
via $\nnbar$ rescattering. 
The quantity $T_{LL'}(p,p';E_k)$ is the $\nnbar$ scattering
amplitude in the coupled $^3S_1$-$^3D_1$ partial wave and is
the solution of a corresponding Lippmann-Schwinger equation:
\begin{eqnarray}
&&T_{L''L'}(p'',p';E_k)=V_{L''L'}(p'',p')+
\sum_{L}\int_0^\infty \frac{dpp^2}{(2\pi)^3} \, V_{L''L}(p'',p)
\frac{1}{2E_k-2E_p+i0^+}T_{LL'}(p,p';E_k)~, 
\label{LS}
\end{eqnarray}
see Ref.~\cite{Kang}. For the potential $V$ in Eq.~(\ref{LS}) we utilize 
a recently published interaction derived within chiral EFT \cite{Kang} and fitted to 
results of a new PWA of $\ppbar$ scattering data \cite{Zhou2012}, and one 
of the phenomenological $\nnbar$ models constructed by the 
J\"ulich group \cite{Hippchen}. In the above equations 
$\sqrt{s} = 2\,E_k=2\sqrt{M^2_p+k^2}$, where $k$ is the $\ppbar$ on-shell 
momentum.

The bare $\nnbar \gamma$ vertex functions, $f^0_L$ ($L$=0,2) in Eq.~(\ref{FSI}),
can be written in terms of bare EMFFs, $G^0_E$ and $G^0_M$, in
complete analogy to Eq.~(\ref{FOFA}). 
On a microscopic level these quantities are given by the direct coupling 
of the photon to the $\nnbar$ system. But they can be also expressed in terms 
of the coupling of the photon to the hadrons through intermediate vector mesons
($\rho$, $\omega$, $\phi$, etc.) which forms the basis of the 
vector meson dominance (VMD) model~\cite{Yan,Lomon,Korner,Williams}. 
There will be also contributions to $f^0_L$ (or, equivalently, to
$G^0_E$ and $G^0_M$) from intermediate mesonic 
states such as $\gamma \to \pi^+\pi^- \to \ppbar$, etc. Thus, 
in principle, $f^0_0$ and $f^0_2$ are complex and can depend on the total energy 
and on the (off-shell) momentum of the $\nnbar$ system. 
 
In the present study we assume that the whole energy dependence
of the dressed vertex functions $f_L$ is generated by the FSI alone and that 
$f^0_0$ and $f^0_2$ themselves are energy-independent. In particular, we 
interpret the explicit 
dependence of $f^0_L$ on $\sqrt{s}$ that is implied by Eq.~(\ref{FOFA}) as a
dependence on the momentum of the $\nnbar$ system. Accordingly, we use
\ba
f^0_0(p) &=& \left(G^0_M + \frac{M_p}{2E_p}\, G^0_E\right) = 
\left(G^0_M + \frac{M_p}{2\sqrt{M_p^2+p^2}}\, G^0_E\right)~,\nonumber \\
f^0_2(p) &=& \frac{1}{\sqrt{2}}\left(G^0_M - \frac{M_p}{E_p}\,G^0_E\right)
= \frac{1}{\sqrt{2}}\left(G^0_M - \frac{M_p}{\sqrt{M_p^2+p^2}}\,G^0_E\right)~,
\label{FOFA0} 
\ea 
for the bare vertex functions, where $p$ is the center-of-mass momentum 
in the $\nnbar$ system, and we assume that $G^0_E$ and $G^0_M$ are real and constant.

The replacement $\sqrt{s} \to 2E_p$ is anyhow required in order to 
guarantee the correct threshold behavior of the {$D$-wave} vertex function
$f^0_2(p)$ which has to behave like $\propto p^2$. 
Indeed, the partial-wave representation of the 
$\eebar \leftrightarrow \ppbar$ amplitudes in form of
Eqs.~(\ref{FOFA}) or (\ref{FOFA0}) is rather instructive
because it makes clear that the condition 
$G^0_E=G^0_M$ and/or $G_E=G_M$ at the $\ppbar$ threshold
is mandatory for implementing the proper threshold 
behavior of the $D$-wave amplitude. 
Assumptions like $|G_E|=0$ imposed in the past in an
analysis of the neutron form factor in the timelike region
for energies fairly close to the threshold \cite{Antonelli}
constitute a drastic violation of this condition. 

Our assumption that $G^0_E$ and $G^0_M$ are 
constant automatically implies that we have to set $G^0_E=G^0_M$. 
$G^0_E$ ($G^0_M$) is taken to be real because any overall 
phase drops out in the evaluation of observables. 
Thus, there is only a {\em single} free parameter in our calculation.
The bare vertex functions $f^0_0$ and $f^0_2$ are calculated 
from Eq.~(\ref{FOFA0}) and inserted into Eq.~(\ref{FSI}). 
Due to the FSI the resulting dressed vertex functions $f_0$ and 
$f_2$ are energy-dependent and also complex.
Inverting Eq.~(\ref{FOFA}) we can obtain $G_E$ and $G_M$ and
then evaluate any $\eebar \leftrightarrow \ppbar$ observable
based on the formulae provided at the beginning of this
section. Note that also $G_E$ and $G_M$ are complex quantities
and, in general, $G_E \neq G_M$ where the difference is likewise 
solely due to the FSI. 

%%%%%%%%%%%%%%%%%%%%%%%%%%%%%%%%%%%%%%%%%%%%%%%%%%%%%%%%%%%%%%%%%%
\section{Results}
\label{sec:res} 

For evaluating the FSI effects we employ amplitudes generated from an 
$\nnbar$ interaction that was recently derived by us within chiral 
EFT \cite{Kang}. In that reference, $\nnbar$ potentials 
up to next-to-next-to-leading order (NNLO) were constructed, based on a modified 
Weinberg power counting, in close analogy to pertinent studies 
of the nucleon-nucleon interaction \cite{Epe05}.
The low-energy constants associated with the arising contact interactions
are fixed by a fit to phase shifts and inelasticities provided by a recently
published phase-shift analysis of $\ppbar$ scattering data \cite{Zhou2012}.
In the $^3S_1$--$^3D_1$ partial wave that is needed for the study of the reaction 
$\ppbar \leftrightarrow \eebar$ good overall agreement with the antinucleon-nucleon 
phase shifts and inelasticities was obtained up to laboratory energies of around 
200 MeV \cite{Kang}. For convenience the corresponding results are reproduced here,
see Fig.~\ref{fig:3SD1}. 
Accordingly, in the present study we restrict ourselves to excess energies 
$Q = \sqrt{s}-2M_p$ of around 
100~MeV in the $\nnbar$ system. In any case, it is primarily the threshold region 
where we expect that FSI effects are relevant and determine the energy dependence 
of the observables. At higher kinetic energies or, generally, over
a larger energy region, the intrinsic energy- and momentum dependence of the 
$\nnbar$ production mechanism itself may become significant or even dominant and
then our assumption that $G^0_E$ and $G^0_M$ are constant is no longer valid.

Besides the EFT interaction we consider again the J\"ulich $N\bar N$ model A(OBE) 
\cite{Hippchen}, which has already been used in our earlier study \cite{Sibirtsev}. 

\begin{figure}[htb!]
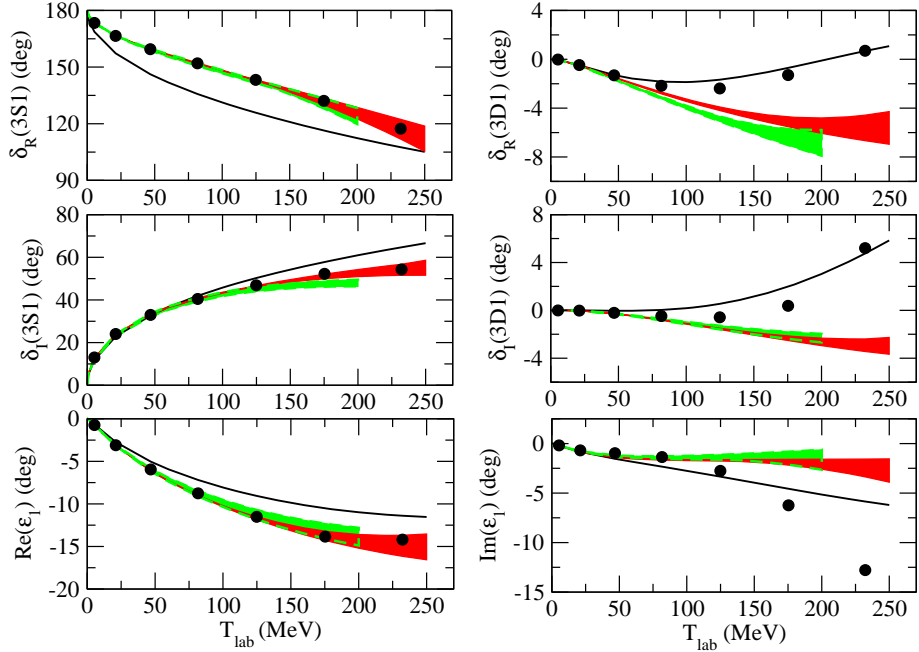

\begin{center}
\includegraphics[width=120mm,clip]{13SD1.eps}
\vskip 0.1cm
\includegraphics[width=120mm,clip]{33SD1.eps}
\caption{Real and imaginary parts of the phase shift in the $^3S_1$--$^3D_1$ partial wave
in the isospin $I=0$ and $I=1$ channels.
The red/dark band shows the chiral EFT results up to NNLO while the green/light band are 
results to NLO. 
The bands reflect the cutoff dependence of the results as discussed in Ref.~\cite{Kang}.
The solid line is the prediction of the J\"ulich $\nnbar$ model A(OBE) \cite{Hippchen}.
The circles represent the solution of the partial-wave analysis of Ref.~\cite{Zhou2012}.
}
\label{fig:3SD1}
\end{center}
\end{figure}

Results for the $\eebar \to \ppbar$ reaction cross section are displayed in
Fig.~\ref{fig:BaBar} as a function of $Q$
and compared with experiments \cite{Aubert0,Lees,DM1,Antonelli}. 
We are interested in the near-threshold region and, therefore, we compare
to the BaBar data with a smaller bin size listed in Table VII of their
papers \cite{Aubert0,Lees}. Since the old and new BaBar data are given 
for precisely the same bins we shifted the 2006 data \cite{Aubert0} to 
slightly higher $Q$ values in Fig.~\ref{fig:BaBar} for a better discrimination.
 
As said above, there is only a single parameter in our calculation, namely $G^0_E$,  
which, in essence, amounts to an overall normalization factor. 
It is fixed by a $\chi^2$ fit to the $\eebar \to \ppbar$ cross section 
data up to $Q \approx 60$~MeV for each of the considered $\nnbar$ interactions.
We want to emphasize again that the energy dependence of the cross section itself
is not influenced by this parameter.
It is given entirely by the FSI effects generated by the various potentials. 
In case of the EFT interactions (at NLO and NNLO) bands are shown. Those
bands reflect the cutoff dependence of the corresponding results 
and can be viewed as an estimate for the theoretical uncertainty of the
interactions, cf. the discussion in Ref.~\cite{Kang}.

Obviously the energy dependence of the $\eebar \to \ppbar$ cross section
is very well reproduced by all $\nnbar$ potentials considered for the FSI,
over the whole energy range up to 100 MeV. This is reflected in the 
achieved $\chi^2/$dof which amounts to 0.81 $\ccc$ 1.01 and 0.63 $\ccc$ 0.71 
for the NLO and NNLO interactions, respectively, and
to 0.64 for the J\"ulich model A(OBE). This is a strong support for the 
conjecture that the energy dependence exhibited by the cross section is 
dominated more or less completely by the one of the $\nnbar$ interaction.
It is interesting to compare the present result with that of our earlier 
study \cite{Sibirtsev}, where only the $^3S_1$ partial wave was taken into
account and which relied on the Migdal-Watson approximation with regard
to the treatment of FSI effects. In that work only the rapid rise of the
cross section close to the threshold could be reproduced and visible
deviations started already at excess energies around 50~MeV. Now, with
the coupling to the $^3D_1$ partial wave included and an accurate treatment
of the FSI effects, there is quantitative agreement with the data (within
the error bars) up to significantly higher energies.

\begin{figure}[t!]
\begin{center}
\includegraphics[width=120mm,clip]{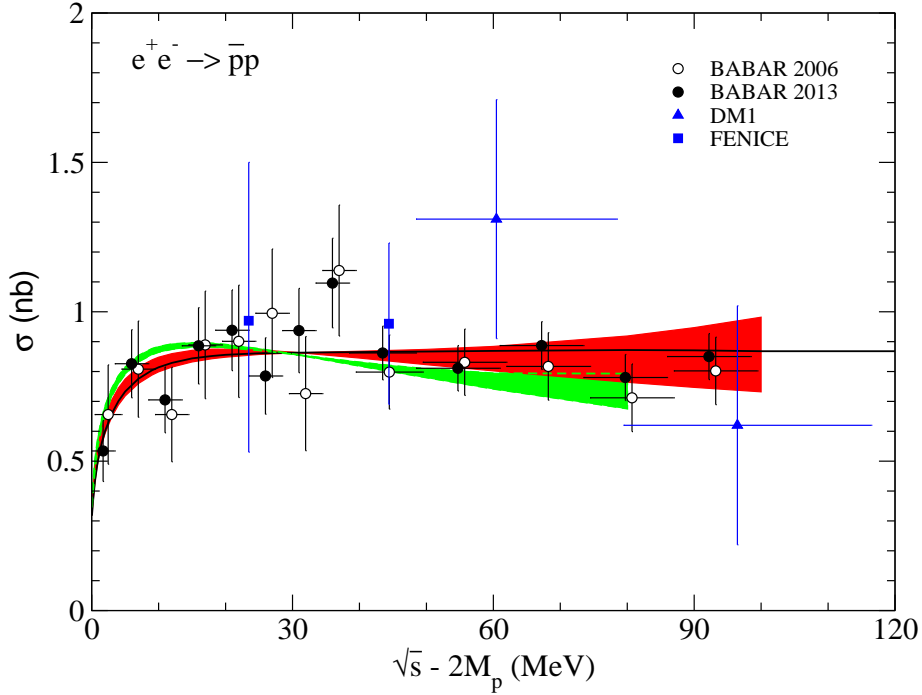}
\caption{Cross section of the reaction $\eebar \to \ppbar$ 
as a function of the excess energy. 
The data are from the DM1~\cite{DM1} (triangles), FENICE~\cite{Antonelli} 
(squares), and BaBar~\cite{Aubert0} (empty circles), \cite{Lees} (filled circles) collaborations. 
The red/dark band shows results based on the $\nnbar$ amplitude of the chiral EFT 
interaction up to NNLO while the green/light band are those for NLO. 
The solid line is the result for the $\nnbar$ amplitude predicted by the
J\"ulich model A(OBE) \cite{Hippchen}.
The BaBar 2006 data are shifted to slightly higher $Q$ values, see text. 
}
\label{fig:BaBar}
\end{center}
\end{figure}

The results in Fig.~\ref{fig:BaBar} and those presented below
are all obtained by using the $\ppbar$ amplitude in Eq.~(\ref{FSI}) 
which is the sum of the isospin $I=0$ and
$I=1$ amplitudes, i.e. $T^{\ppbar} = (T^{I=1}+T^{I=0})/2$. However,
we did perform exploratory calculations employing also $T^{I=1}$ and $T^{I=0}$
separately. The corresponding results turned out to be very similar
to each other and also to the one based on the $\ppbar$ amplitude. Indeed, in all 
cases we obtain excellent agreement with the energy dependence exhibited 
by the data. Thus, we do not see any evidence for a possible dominance of the 
isoscalar amplitude as suggested in Ref.~\cite{Dmitriev13}. 

A comparison with data for the inverse reaction, $\ppbar \to \eebar$, 
that were taken by the PS170 Collaboration at LEAR is provided in 
Fig.~\ref{fig:PS170}. This cross section is related to the one for
$\eebar \to \ppbar$ by detailed balance and time-reversal invariance, 
i.e. by
\begin{eqnarray}
\sigma_{\ppbar \to \eebar} \simeq
\frac{k_e^2}{k_p^2} \, \sigma_{\eebar \to \ppbar} \ .
\label{det}
\end{eqnarray}
There is a well-known systematical difference between the $\eebar \to \ppbar$ 
and $\ppbar \to \eebar$ cross section data \cite{Denig}, 
where the latter are smaller by a factor of about 1.47.   
But once we take that into account by a proper renormalization 
of our results (using the same renormalization factor for all considered $\nnbar$
interactions) we reproduce the PS170 measurement rather nicely
as can be seen in Fig.~\ref{fig:PS170}. Obviously, the energy
dependence of the $\ppbar \to \eebar$ cross section revealed by 
the PS170 data \cite{Bardin} is perfectly consistent with the one
of the $\eebar \to \ppbar$ data measured by the BaBar collaboration 
\cite{Aubert0,Lees}. 

\begin{figure}[tb!]
\begin{center}
\includegraphics[width=120mm,clip]{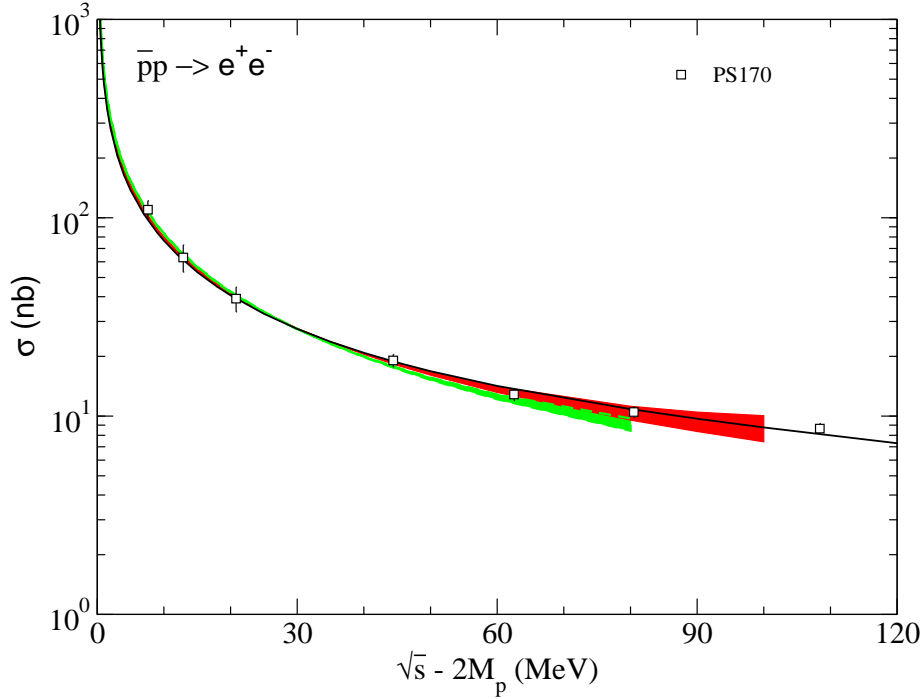}
\caption{Cross section of the reaction $\ppbar \to \eebar$ 
as a function of the excess energy. 
The data are from the PS170~\cite{Bardin} collaborations. 
Same description of curves as in Fig.~\ref{fig:BaBar}. 
}
\label{fig:PS170}
\end{center}
\end{figure}

Results for the effective proton form factor in the timelike region, defined in
Eq.~(\ref{eq:Geff}), are displayed in Fig.~\ref{fig:peff}. Data for this
quantity, which provides a quantitative indication for the deviation of the 
measured cross section from the point-like case \cite{Denig} can be readily 
found in those publications where experiments for 
$\eebar\leftrightarrow\ppbar$ were reported~\cite{Aubert0,Lees,DM1,Antonelli}. 
The effective form factor for the point-like case would be simply a straight 
line in Fig.~\ref{fig:peff}, i.e. there would be no dependence on the excess energy. 
The experimental form factor, on the other hand, shows a significant rise for 
energies close to the threshold as already mentioned in the Introduction. 
Our results that include the $\nnbar$ FSI are very well in line with this
behaviour. This is not surprising in view of the fact that we
reproduce the $\eebar \leftrightarrow \ppbar$ cross sections that form the 
basis for determining the effective proton form factor, see Eq.~(\ref{eq:Geff}). 

\begin{figure}[htb!]
\begin{center}
\includegraphics[width=120mm,clip]{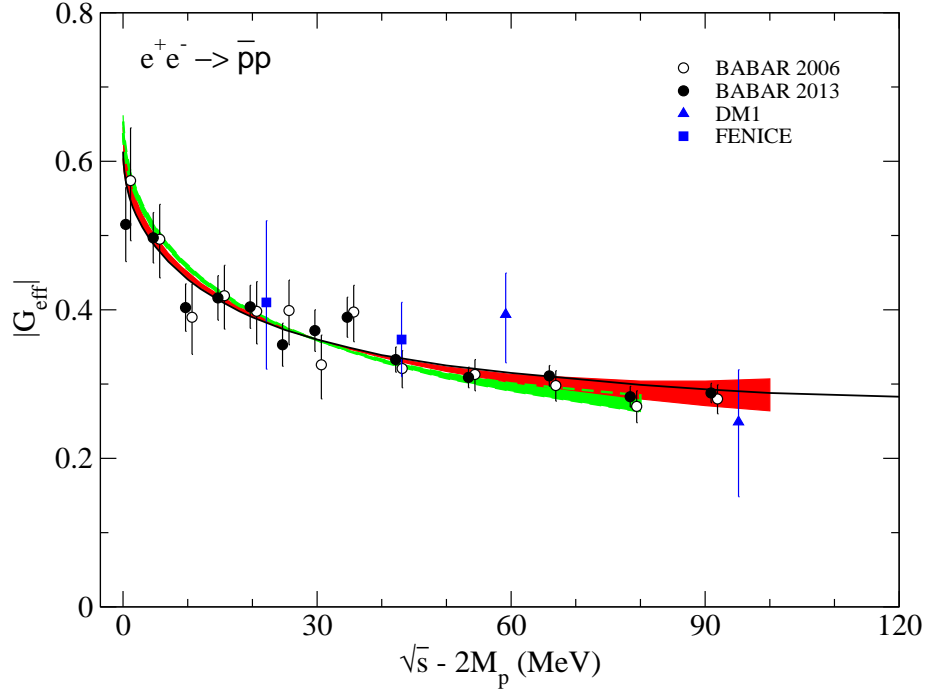}
\caption{Effective proton form factor, defined in Eq.~(\ref{eq:Geff}),
as a function of the excess energy. 
The data are from the DM1~\cite{DM1} (triangles), FENICE~\cite{Antonelli} 
(squares), and BaBar~\cite{Aubert0} (empty circles), \cite{Lees} (filled squares) collaborations. 
Same description of curves as in Fig.~\ref{fig:BaBar}. 
The BaBar 2006 data are shifted to slightly higher $Q$ values, see text. 
}
\label{fig:peff}
\end{center}
\end{figure}

There is also experimental information on angular distributions. 
For the reaction $\eebar \to \ppbar$ such distributions are
provided for different intervals of the $\ppbar$ invariant mass
\cite{Aubert0,Lees}. We consider here solely the lowest two,
because only those concern the energy region for which our 
EFT $\nnbar$ potentials are designed. 
The corresponding intervals in terms of the excess energies are 
$0 \leq Q \leq 73$ MeV and $73 \leq Q \leq 148$ MeV.
It is clear that data which sample over such a 
large energy range cannot reflect any more subtle variations of the 
angular distribution with energy. Thus, we perform our calculations 
for the average energies of those intervals, namely $Q=$ 36.5 MeV and 
110.5 MeV. The results are confronted with the BaBar data in
Fig.~\ref{fig:BaBard}. There is a remarkable agreement in case of
EFT interactions. We want to
emphasize that the angular distributions are genuine predictions.
They are completely fixed by the properties of the employed $\nnbar$ FSI.
Note that the overall normalization is arbitrary because only
the number of events are given in Refs.~\cite{Aubert0,Lees}. 
Again the 2006 data \cite{Aubert0} are slightly shifted for a better 
discrimination.

\begin{figure}[htb!]
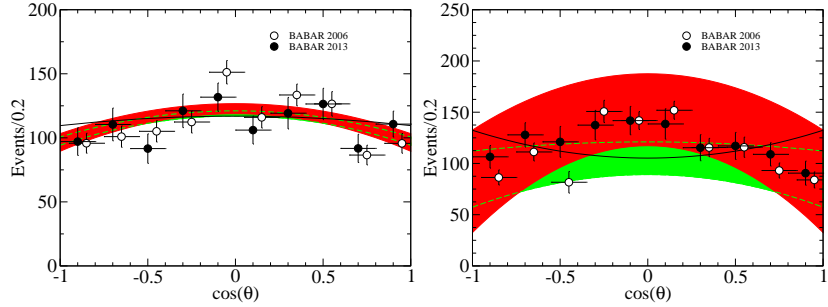

\begin{center}
\includegraphics[height=40mm,clip]{diffBaBar1.eps}
\includegraphics[height=40mm,clip]{diffBaBar3.eps}
\caption{Differential cross section for $\eebar \to \ppbar$ at
the excess energies $Q = 36.5$ MeV (left) and 
$Q = 110.5$ MeV (right).  
The data are an average
over $0 \leq Q \leq 73$ MeV and over $73 \leq Q \leq 148$ MeV,
respectively, and are taken from Refs.~\cite{Aubert0,Lees}. 
Same description of curves as in Fig.~\ref{fig:BaBar}. 
The BaBar 2006 data are slightly shifted, see text. 
}
\label{fig:BaBard}
\end{center}
\end{figure}

In case of $\ppbar \to \eebar$ proper differential cross sections 
were measured, at laboratory momenta of 416, 505, 581, 681, and 888 MeV/c 
\cite{Bardin}. Also here we restrict ourselves to energies 
within the range where our EFT interactions are applicable which means 
we compare our results to the data at the first four momenta only. 
The corresponding excess energies are 43.5, 62.6, 80.9 and
107.5 MeV, respectively, and pertinent results are presented in
Fig.~\ref{fig:PS170d}. Again there is reasonable agreement of the
results based on the EFT interactions with the trend exhibited
by the experiment. 
 
Note that in both cases the highest considered energy, 
$Q \approx 110$ MeV ($T_{lab}\approx 220$ MeV), is already in a region 
where our NLO and NNLO interactions no longer reproduce the $\ppbar$ 
amplitudes of the PWA sufficiently well, see Fig.~\ref{fig:3SD1}. 
Thus, those results may be questionable and they are also afflicted 
by large uncertainties as reflected by the bands.
We show them only for illustrative purposes. 

The prediction based on the phenomenological J\"ulich model disagrees 
with the trend shown by the BaBar data at the higher energy but is
still in line with the PS170 measurement at practically the same
excess energy ($Q\approx 107$~MeV). 
This $\nnbar$ potential produces a different $D$ wave admixture 
in the $\eebar \leftrightarrow \ppbar$ amplitude as compared to the 
EFT interactions -- which is not surprising in view of the differences 
in the corresponding $\nnbar$ phase shifts, cf. Fig.~\ref{fig:3SD1}. 
Obviously, the differential cross sections are more sensitive to 
details of the $\nnbar$ interaction than the (energy dependence of the)
integrated cross section where the results for all $\nnbar$
considered interactions more or less coincide. Thus, it would be 
indeed very valuable to have further data on differential cross 
sections with improved statistics. 

\begin{figure}[tb!]
\begin{center}
\includegraphics[width=120mm,clip]{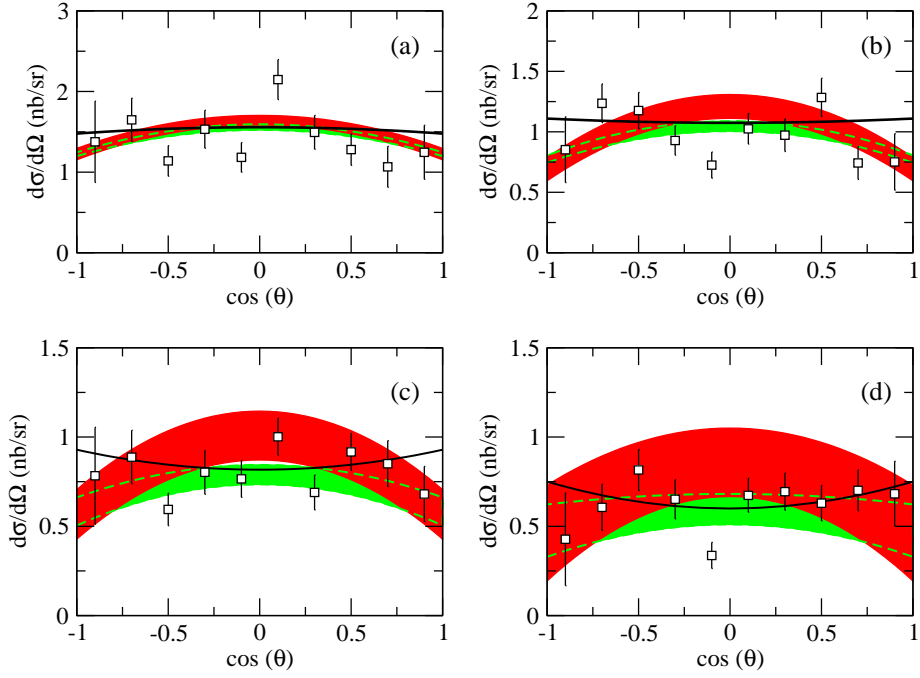}
\caption{Differential cross section for $\ppbar \to \eebar$ at
the excess energies $Q=$ 43.5 (a), 62.6 (b), 80.9 (c), and 107.5 MeV (d), 
respectively. Data are taken from Ref.~\cite{Bardin}. 
Same description of curves as in Fig.~\ref{fig:BaBar}. 
}
\label{fig:PS170d}
\end{center}
\end{figure}

Since our calculation agrees rather well with all measured 
$\eebar \leftrightarrow \ppbar$ observables in the near-threshold
region it is instructive to consider now predictions for other 
quantities like spin observables and also for the EMFFs
$G_E$ and $G_M$ themselves. Results for the latter are presented
in Fig.~\ref{fig:GEGM} where we display the 
modulus and the argument of the ratio $G_E/G_M$ as a function of the 
excess energy. 
The ratio $|G_E/G_M|$ drops to values slightly below 1 right above
the $\ppbar$ threshold but quickly turns to values larger than 1 
with increasing energy. At higher energies the EFT interaction fitted 
to the $\nnbar$ PWA and the J\"ulich meson-exchange model exhibit 
different trends for the ratio. Again this is simply due to differences 
in the pertinent $\nnbar$ amplitudes at these energies, as reflected in 
the phase shifts shown in Fig.~\ref{fig:3SD1}. In that figure one
can also see that the EFT interaction does not reproduce the $^3D_1$
phase shifts of the $\nnbar$ PWA so well anymore for energies above
$T_{lab}\approx$ 130 MeV ($Q \approx$ 65 MeV). Thus, since 
the $D$-waves are responsible for the deviation of $|G_E/G_M|$ from 1, 
one should refrain from associating the results based on our EFT interaction 
with those implied by the original $\nnbar$ amplitudes 
of the PWA at higher energies. 
In any case, there is also an increasing uncertainty due to the
cutoff dependence as visible from the bands. 

\begin{figure}[htb!]
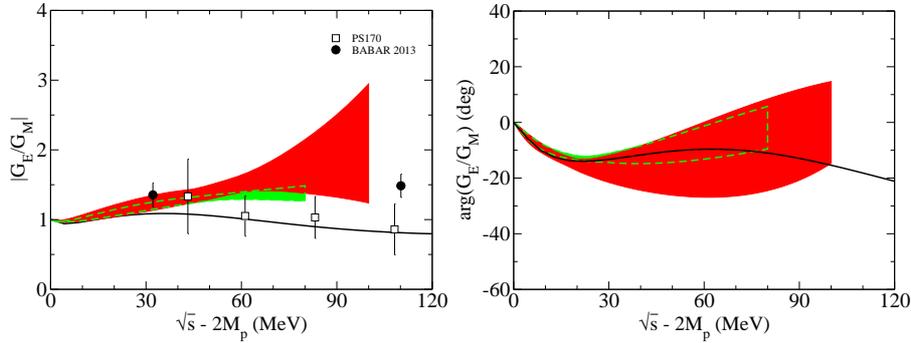

\begin{center}
\includegraphics[height=45mm,clip]{ratio.eps}
\includegraphics[height=45mm,clip]{FFphase.eps}
\caption{$|G_E/G_M|$ and $\rm{arg}(G_E/G_M)$
as a function of the excess energy. 
Data are taken from Refs.~\cite{Bardin} and \cite{Lees}. 
Same description of curves as in Fig.~\ref{fig:BaBar}. 
}
\label{fig:GEGM}
\end{center}
\end{figure}

Predictions for the phase between $G_E$ and $G_M$ are shown in Fig.~\ref{fig:GEGM}. 
It is negative over a larger energy range starting from the threshold. Also
here the EFT interaction and the J\"ulich model exhibit a different behavior 
for higher energies. Overall, the phase remains small with values between 
$\pm$20 degrees. 

Finally, let us present some results for spin-dependent observables for the 
reaction $\ppbar \to \eebar$, in particular, for the analyzing power $A_y$ and the 
spin-correlation parameters $A_{ij}$. These observables can be
written in terms of the ($\eebar \to \ppbar$) helicity amplitudes given in Eq.~(\ref{hel1}) 
following the standard procedure outlined in Refs.~\cite{Holzenkamp,Bystricky}:
\begin{eqnarray}
\nonumber
A_y &=& -({\rm Im}\,\phi_5^* (\phi_3-\phi_4)
-{\rm Im}\,\phi_6^* (\phi_1+\phi_2))/ D , \\
\nonumber
A_{xx} &=& ({\rm Re}\,[\phi_1^* \phi_2+\phi_3^*\phi_4] +|\phi_5|^2-|\phi_6|^2) / D , \\
\nonumber
A_{yy} &=& ({\rm Re}\,[\phi_1^* \phi_2-\phi_3^*\phi_4] +|\phi_5|^2+|\phi_6|^2) / D , \\
\nonumber
A_{zz} &=&-(|\phi_1|^2+|\phi_2|^2-|\phi_3|^2-|\phi_4|^2+2|\phi_5|^2-2|\phi_6|^2) /(2\,D) , \\
A_{xz} &=&-({\rm Re}\,\phi_5^* (\phi_3-\phi_4)
+{\rm Re}\,\phi_6^* (\phi_1+\phi_2))/ D , 
\label{Pol}
\end{eqnarray}
where $D = (\sum_{i=1}^8 |\phi_i|^2)/ 2$.
Corresponding expressions in terms of $G_E$ and $G_M$ can be
found in Refs.~\cite{GaT05,Buttimore31,Bilenky93,GaT11}.
Our predictions for $A_y$ and $A_{ij}$ at the excess energy $Q=45$ MeV 
are depicted in Figs.~\ref{fig:Ay} and \ref{fig:Spin1}, respectively. 
These observables show clear symmetry properties in case of the
one-photon exchange approximation considered here, as one can
read off the formulae given in Ref.~\cite{Buttimore31}. 
Specifically, $A_y$ and $A_{xz}$ are proportional to $\sin 2\theta$, 
and $A_{xx}$ and $A_{yy}$ are proportional to $\sin^2\theta$. 
The magnitudes of $A_y$ and $A_{xz}$ are given by the relative
phase of $G_E$ and $G_M$, namely by Re\,$(G_E^{} G^*_M)$ in case of the
former and by Im\,$(G_E^{} G^*_M)$ for the latter \cite{Buttimore31}. 
Predictions for these quantities can be found in Fig.~\ref{fig:GEGM2},
again as a function of the excess energy.

\begin{figure}[t!]
\begin{center}

\vspace{4mm}

\includegraphics[height=80mm,clip]{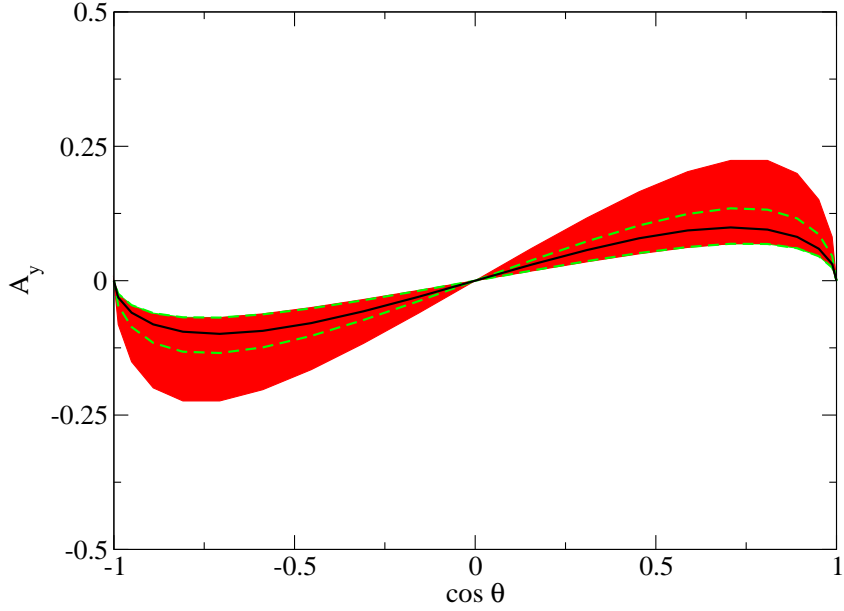}
\caption{Analyzing power for $\ppbar \to \eebar$ at the excess energy
$Q=45$ MeV.
Same description of curves as in Fig.~\ref{fig:BaBar}. 
}
\label{fig:Ay}
\end{center}
\end{figure}

\begin{figure}[htb!]
\begin{center}
\includegraphics[height=80mm,clip]{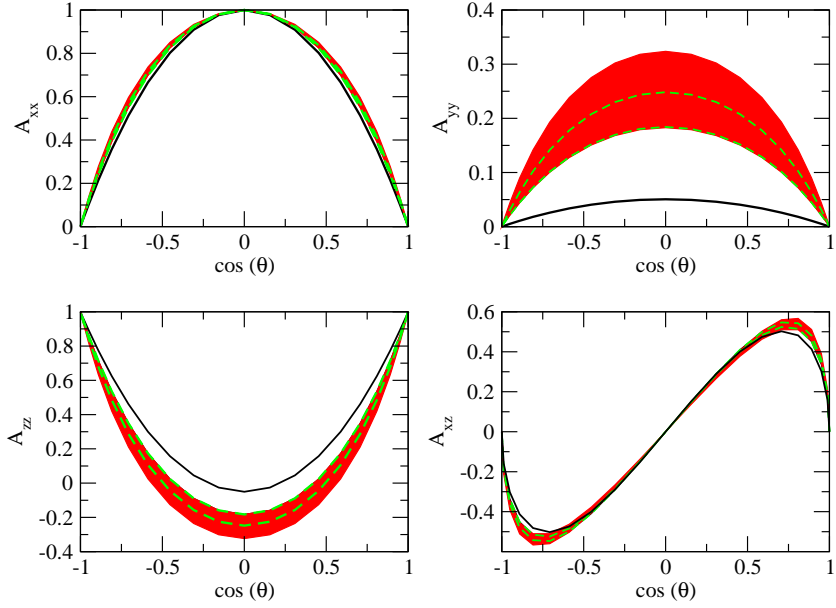}
\caption{Spin correlation parameters for $\ppbar \to \eebar$ at the excess energy
$Q=45$ MeV.
Same description of curves as in Fig.~\ref{fig:BaBar}. 
}
\label{fig:Spin1}
\end{center}
\end{figure}

\begin{figure}[htb!]
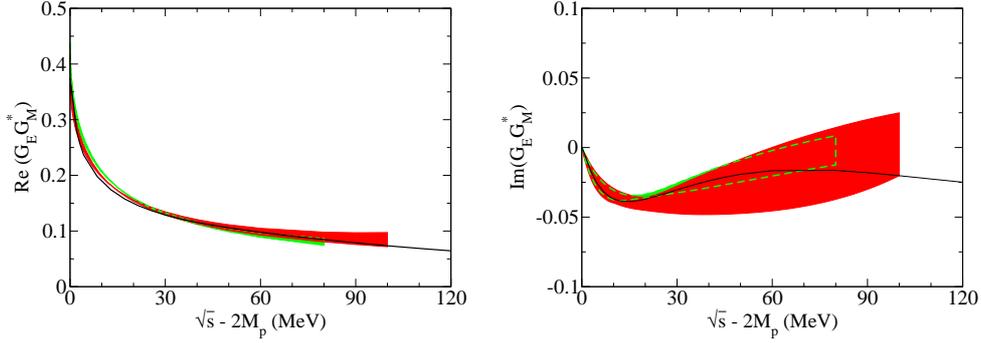

\begin{center}
\includegraphics[height=45mm]{ReGEGM.eps}{$\phantom{xx}$} 
\includegraphics[height=45mm]{ImGEGM.eps}
\caption{Re\,$(G_E^{} G^*_M)$ and Im\,$(G_E^{}G^*_M)$
as a function of the excess energy. 
Same description of curves as in Fig.~\ref{fig:BaBar}. 
}
\label{fig:GEGM2}
\end{center}
\end{figure}

Results for spin-dependent observables have been also published
by other authors \cite{Dubnickova,Dmitriev11,Bilenky93,GaT11,GaT12}
based on various models, however, in general, for much higher
energies. 

An issue that arises in the context of any observed enhancement
in the near-threshold $\ppbar$ production cross sections or in 
the corresponding $\ppbar$ invariant mass spectra is the question
whether this is a signal for an $\nnbar$ bound state. 
Indeed sometimes it is argued that explanations in terms of FSI effects 
or via an $\nnbar$ bound state would mutually exclude each other. 
This is clearly not the case
as we know very well from studies of near-threshold pion production 
in the reaction $NN\to NN\pi$ \cite{Hanhart}. In this case the $NN$ forces in the 
$^1S_0$ and/or $^3S_1$ final $NN$ state allow one to achieve a quantitative 
description of the enhancements seen in the measurements and the very same 
forces also produce the deuteron bound state in the $^3S_1$--$^3D_1$
partial wave and a virtual state in the $^1S_0$.  
Of course, not every enhancement seen in the experiments is a signal for 
forces that are strong enough to produce a pole in the near-threshold
region. For example, a pronounced near-threshold enhancement was
also observed in the $\Lambda p$ invariant mass spectrum as measured 
in the reaction $pp\to p\Lambda K^+$, see for example \cite{Roeder}. 
However, evidently there is no near-threshold $\Lambda p$ bound state. 
Thus, one has to be cautious with conclusions concerning the
existence of such bound states from production reactions.  

Anyway, let us come back to the $\nnbar$ interaction investigated here. 
For the employed EFT potentials a search for poles near
the threshold was performed and the results were reported in 
Ref.~\cite{Kang}. No bound state was found for the 
$^{3}S_1$--$^{3}D_1$ partial wave in the isospin $I=1$ channel. 
There is a pole in the $I=0$ channel, however, it corresponds to
a ``binding'' energy of $Q_0 = +(5.6\ccc 7.7)-{\rm i}\,(49.2\ccc 60.5)$ MeV,
depending on the cutoffs, at NLO and
$Q_0= +(4.8\ccc 21.3)-{\rm i}\,(60.6\ccc 74.9)$ MeV at NNLO \cite{Kang}. 
We used quotation marks above because  
the positive sign of the real part of $Q_0$ indicates that these poles
are actually located above the $\bar NN$ threshold. They lie on the 
physical sheet and, therefore, do not correspond to resonances either.
In Ref.~\cite{Badalyan82} such poles are referred to as 
unstable bound states. 

%%%%%%%%%%%%%%%%%%%%%%%%% corrections end here %%%%%%%%%%%%%%%%%%%%%%%%%%
\section{Conclusions}

We analyzed the reactions $\ppbar \to \eebar$ and $\eebar \to \ppbar$ 
in the near-threshold region with specific emphasis on the role played 
by the interaction in the initial- or final $\nnbar$ state. The
study is based on the one-photon approximation for the elementary
reaction mechanism, but takes into account rigorously the effects of 
the $\ppbar$ interaction. For the latter we utilized a recently 
published $\nnbar$ potential derived within chiral effective field theory 
\cite{Kang} and fitted to results 
of a new partial-wave analysis of $\ppbar$ scattering data \cite{Zhou2012}, 
and also one of the phenomenological $\nnbar$
meson-exchange models constructed by the J\"ulich group \cite{Hippchen}. 

Our results confirm the conjecture drawn from previous studies
\cite{Sibirtsev,Dmitriev07,Chen08,Chen10,Dalkarov,Dmitriev11,Dmitriev13,Dmitriev13a}
that the pronounced energy dependence of the
$\eebar \leftrightarrow \ppbar$ cross section, seen in pertinent 
experiments, is indeed primarily due to the $\ppbar$ interaction. 
However, the evidence provided now is much more convincing. First the
present calculation is technically superior to the earlier ones because
it relies on an rigorous treatment of the FSI effects.
Secondly, it utilizes $\nnbar$ 
amplitudes that have been determined from a PWA. And, finally,  
by including not only the $^3S_1$ but also the $^3D_1$ partial wave
the energy dependence of the experimental cross sections can be 
described quantitatively and over a significantly larger energy region. 
In addition, even existing data on angular distributions are well reproduced. 

Based on our results for the reactions $\eebar \leftrightarrow \ppbar$ 
we can produce reliable predictions for the proton electromagnetic form
factors $G_E$ and $G_M$ in the timelike region, for $q^2$ near the
$\nnbar$ threshold. 
The effective proton form factor usually considered in the literature 
exhibits a strong $q^2$-dependence for $q^2 \approx (2 M_p)^2$ 
and this behavior is perfectly described by our calculation. 
The strong $q^2$-dependence is likewise a consequence of the 
interaction in the $\ppbar$ system. 
For the ratio $|G_E/G_M|$ we predict a non-trivial energy dependence. 
The ratio drops to values slightly below 1 right above the $\nnbar$ threshold 
but turns to values larger than 1 within a couple of MeV. The phase between
the form factors, arg$(G_E/G_M)$, is negative for energies close to the
$\nnbar$ threshold with values in the order of $-10$ to $-20$ degrees. 

The predictions for the differential cross sections, and also for 
$|G_E/G_M|$ and arg$(G_E/G_M)$, based on the chiral EFT interaction and
on the phenomenological J\"ulich $\nnbar$ potential, show different tendencies
with increasing energy. The presently available data (for the differential
cross section) are afflicted with sizable uncertainties and, thus, do not 
allow to discriminate between these differences. Moreover, the BaBar and 
the PS170 data themselves seem to
be incompatible at higher excess energies as visible, for example, in the 
extracted ratio $|G_E/G_M|$ \cite{Lees}, see also Fig.~\ref{fig:GEGM}. 
Therefore, it would be very interesting to perform new measurements of the reactions
$\eebar \leftrightarrow \ppbar$ with improved statistics. As discussed
in the review \cite{Denig}, such experiments could be accomplished at the
VEPP-2000 accelerator in Novosibirsk \cite{Solodov} or the 
BEPC-II collider in Beijing (for $\eebar \to \ppbar$),
but also by the $\bar {\rm P}$ANDA set-up at the planned FAIR facility 
in Darmstadt \cite{Sudol} (for the inverse reaction $\ppbar \to \eebar$). 
Evidently, aside from pinning down the electromagnetic form factors in
the time like region more accurately, such data would also provide further
constraints on our knowledge of the elementary $\nnbar$ interaction 
where direct information in the near-threshold region is still rather
scarce. 

\section*{Acknowledgements}

We would like to thank Kanzo Nakayama for clarifying discussions. 
This work is supported in part by the DFG and the NSFC through
funds provided to the Sino-German CRC 110 ``Symmetries and
the Emergence of Structure in QCD'' and by the EU Integrated
Infrastructure Initiative HadronPhysics3.

\end{document}